\documentclass[%
mathleft,%
final,%
]{an}
\usepackage{graphicx}
\usepackage{times}
\usepackage{natbib}
\bibpunct{(}{)}{;}{a}{}{,}
\usepackage{url}
\newcommand{\kms}{km\,s$^{-1}$}

\newcommand{\COBOLD}{{\sf CO$^5$BOLD}}
\newcommand{\cobold}{\COBOLD}
\newcommand{\linfor}{{\sf Linfor3D}}
\def\araa{ARA\&A}%
\def\apj{ApJ}%
\def\apjl{ApJ}%
\def\aap{A\&A}%
\def\aaps{A\&AS}%
\def\mnras{MNRAS}%
\def\pasp{PASP}%
\def\solphys{Sol.~Phys.}%
\def\nat{Nature}%
\begin{document}

\Pagespan{1}{}
\Yearpublication{2014}%
\Yearsubmission{2013}%
\Month{1}%
\Volume{335}%
\Issue{1}%
\DOI{This.is/not.aDOI}%

\title{Isotope Spectroscopy
}

\author{E. Caffau\inst{1,2}\fnmsep\thanks{Corresponding author.
  \email{ecaffau@lsw.uni-heidelberg.de}}
\and  M. Steffen\inst{3,2}
\and  P. Bonifacio\inst{2}
\and  H.-G. Ludwig\inst{1,2}
\and  L. Monaco\inst{4}
\and  G. Lo Curto\inst{5}
\and  I. Kamp\inst{6}
}
\titlerunning{Isotope Spectroscopy}
\authorrunning{E. Caffau et al.}
\institute{
Zentrum f\"ur Astronomie der Universit\"at Heidelberg, Landessternwarte, 
K\"onigstuhl 12, 69117 Heidelberg, Germany
\and
GEPI, Observatoire de Paris, CNRS, Universit\'e Paris Diderot, Place
Jules Janssen, 92190
Meudon, France
\and
Leibniz-Institute for Astrophysics Potsdam (AIP), An der Sternwarte 16,
D-14482 Potsdam, Germany
\and
European Southern Observatory, Casilla 19001, Santiago, Chile
\and
European Southern Observatory, Karl-Schwarzschild-Str. 2, 85748 Garching bei M\"unchen, Germany
\and
Kapteyn Astronomical Institute,
University of Groningen, Postbus 800, 9700 AV Groningen, The Netherlands
}

\received{XXXX}
\accepted{XXXX}
\publonline{XXXX}


\abstract{
The measurement of isotopic ratios provides a privileged
insight both into nucleosynthesis and into the mechanisms
operating in stellar envelopes, such as gravitational settling.
In this article, we give a few examples of how isotopic ratios
can be determined from high-resolution, high-quality stellar spectra.
We consider examples of the lightest elements, H and He, for which the isotopic 
shifts are very large and easily measurable, and examples of heavier
elements for which the determination of isotopic ratios is more difficult.
The presence of $^6$Li in the stellar atmospheres causes a subtle extra
depression in the red wing of the $^7$Li 670.7\,nm doublet 
which can only be detected in spectra of the highest quality. But even with
the best spectra, the derived $^6$Li abundance can only be as good as the
synthetic spectra used for their interpretation. It is now known that
3D non-LTE modelling of the lithium spectral line profiles is necessary 
to account properly for the intrinsic line asymmetry, which is produced by 
convective flows in the atmospheres of cool stars, and can mimic the 
presence of $^6$Li. We also discuss briefly the case of the carbon isotopic 
ratio in metal-poor stars, and provide a new determination of the nickel 
isotopic ratios in the solar atmosphere.
}

\maketitle

\section{Introduction}

The knowledge of the \emph{chemical abundance} pattern of a star gives 
us insights into the previous generations of massive stars that enriched
the interstellar gas. The analysis of the \emph{isotopic ratios} provides us 
with details about the different nuclear reactions involved and their relative 
contributions.
These nuclear reactions could have taken place:
(i) in the interiors of the previous generations of massive stars;
(ii) during the final stages of stellar evolution in supernova explosions;
(iii) in dynamical processes in late evolutionary phases of stars (s-process 
in AGB stars, mixing of CNO-processed material in RGB and AGB stars, hot-bottom 
burning, cool-bottom processing etc.);
(iv) in the interstellar medium due to cosmic ray spallation events.

The analysis of isotopic ratios is considerably more difficult than the 
abundance determination, generally requiring extremely clean, high-quality
spectra (highest spectral resolution and signal-to-noise ratio). 
Even then the isotopic ratio can be 
determined for only about 10\% of the elements for which the photospheric 
abundance can be derived.

In the case that atomic lines are used in the isotopic analysis,
the lighter the element the larger the wavelength shift
between the lines of the different isotopes: for light elements,
the isotopic shift is mainly due to the different masses of the
isotopes, $m$ and $m'$, such that
$\Delta\lambda/\lambda^2 \sim 1/m-1/m'$. 
This effect makes the isotopic analysis for lightest elements easier.
\citet{spite83} analysed the spectra of three metal-poor dwarfs
to derive the deuterium-to-hydrogen ratio. They could not detect the 
component due to deuterium (D) in the blue wing of H$\alpha$, although 
the wavelength separation  between the two components is large enough 
(of the order of 0.2\,nm or 90\,\kms) to easily distinguish both components.
We will discuss in separate sections the other light elements:
He in Sect.\,\ref{helium} and Li in Sect.\,\ref{lithium}.

The isotopic ratio of carbon and oxygen is usually derived from 
molecular lines. \citet{ayres13} derived the isotopic ratios of 
carbon and oxygen by the analysis of infrared molecular lines of CO.
In the solar spectrum, the CO lines are isolated, and, after stacking 
them in a suitable way, the authors  could derive the 
$^{12}$C/$^{13}$C ratio, as well as the ratios of $^{16}$O, $^{17}$O, 
and $^{18}$O in the solar photosphere. In Sect.\,\ref{carbon} we 
investigate the case of the carbon isotopic ratio in metal-poor stars.

For heavy elements, the isotopic shifts are only a fraction of the 
line width, such that only the full width at half maximum 
of the line can be used to derive the isotopic ratio.
For the case of Ba \citep[see e.g.][]{gallagher12},
the hyperfine splitting  (HFS) of the odd isotopes causes a desaturation
of the line and an increase of the width of the feature. The larger the
HFS for the odd isotopes, the easier the determination of the isotopic 
ratios.

For intermediate-mass elements, both the full width at half maximum
and the shape of the line must be taken into account in the analysis.
We illustrate the situation for the case of the solar nickel isotopes
in Sect.\,\ref{nickelsec}.

This paper is not intended to be a review on isotopic 
ratio analysis, but rather presents a few representative examples, mostly
our own results not published elsewhere.

\section{The helium isotopic ratio in Feige\,86}\label{helium}

Feige\,86 (BD\,+30\,2431) is a chemically peculiar halo horizontal-branch
star \citep[][and references therein]{Bonifacio}. 
Its most remarkable peculiarities are a large overabundance
of phosphorus, an underabundance of helium (``He-weak star'') and
a large $^3$He/$^4$He isotopic ratio. The \ion{He}{i} 
2$^1$P--3$^1$D line at 667.8\,nm
has an isotopic shift of 0.05\,nm (22\,\kms), while the corresponding
triplet line at 587.5\,nm   has a negligible isotopic shift.
\citet{Hartoog} observed the 667.8\,nm line in
Feige\,86 at the Lick Observatory
with the Shane 3.0\,m telescope and an image intensifier
that provided a dispersion of 11\AA mm$^{-1}$, corresponding
to a resolving power   R$\sim 15\,000$, too low to resolve the 
two components. However, based on the analysis of the first
moment of the line, \citet{HC} claimed a detection
of $^3$He and an isotopic ratio $^3$He/$^4$He =0.7.
The detection was confirmed by higher resolution,
albeit lower S/N ratio, spectra obtained by
\citet{Bonifacio} with AURELIE at the 
Observatoire de Haute Provence  1.5\,m telescope.
From figure 7 of \citet{Bonifacio}, it could already be
guessed that the isotopic ratio of Hartoog was
overestimated. 

\begin{figure}
\mbox{\includegraphics[width=\linewidth,clip=true]{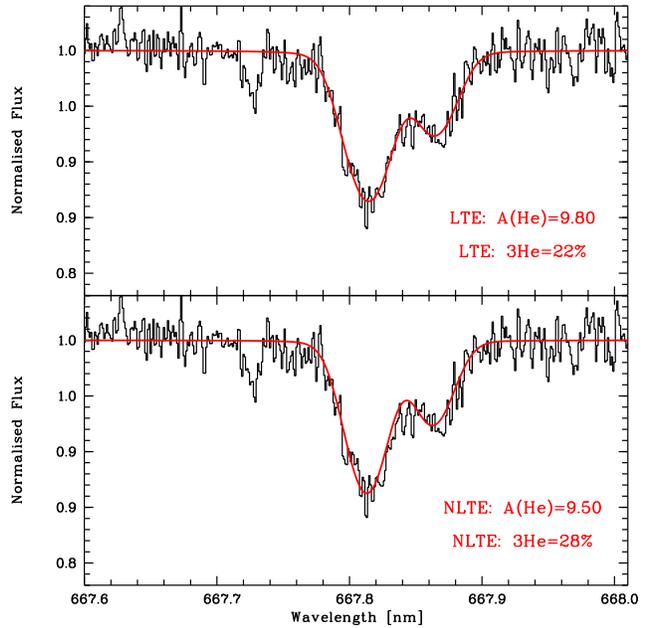}}
\caption{The best fit (solid red) of the helium 667.8\,nm line superimposed 
on the observed HARPS spectrum of Feige\,86 (solid black), both for
LTE (top) and non-LTE (bottom) line formation.}
\label{He_fitiso}
\end{figure}

We observed Feige\,86 with HARPS at the
ESO 3.6\,m telescope on the nights of March 9 and
10, 2006. Each of the two exposures was 1800s long.
The spectra do not show any relative shift after
barycentric correction  and were co-added. The S/N ratio
of the co-added spectrum
around the 667.8\,nm \ion{He}{i} line  is about 70.
The resolving power of HARPS is about 110\,000, and the $^3$He 
line is  clearly resolved (see Fig.\,\ref{He_fitiso}).
This high resolution spectrum shows clearly that Feige\,86
is very sharp-lined, and we estimate a rotational velocity
of $v\,\sin i \approx 2.5$\,kms$^{-1}$ from fits to isolated 
\ion{P}{ii} lines. This is much lower than 
the 11\,kms$^{-1}$ estimated by \citet{SC92}
from the ultraviolet spectra observed with the IUE satellite.
It is likely that in the \citet{SC92} measurement the limiting factor
is the lower resolution of the IUE spectra.
The measured radial velocity is $-28\pm 1$\,kms$^{-1}$, 
in good agreement with lower precision earlier measurements
($-23\pm 1$\,kms$^{-1}$ by \citealt{berger}, and $-21\pm 5$\,kms$^{-1}$ 
by \citealt{PB}).  The star has probably 
a constant radial velocity. It may nevertheless be interesting to
make further radial velocity observations for this star, since, taken at
face value, the different available measurements could signal a long 
period orbital motion. 
 
To determine the helium isotopic ratio
we used spectrum synthesis.
We computed an ATLAS\,12 model atmosphere \citep{kurucz05,Castelli}
with the atmospheric parameters of 16430/4.2/0.0 
\cite[${\rm T}_{\rm eff}$/logg/metallicity,][]{Bonifacio}
and we changed the abundances of He, C, N, O, Mg, Si, P, S, 
Ca, and Fe according to Table\,9 of \citet{Bonifacio} to match 
the chemical composition of Feige\,86.
This model is practically indistinguishable from
the corresponding ATLAS\,9 model for $\tau_{Ross}>-4$.
To synthesise both lines in local thermodynamical equilibrium (LTE) and non-LTE we used the
Kiel code \citep{SH} and  the atomic data in Table\,\ref{atom}.
For the non-LTE computation we used a model atom
consisting of 36 levels of \ion{He}{i} and a \ion{He}{II}
continuum, which should be adequate for A- and B-type stars. 
Collisions with neutral hydrogen were 
treated with the Drawin formalism \citep{Drawin}
as generalised by \citet{SH}, adopting an
enhancement factor $S_{\rm H}=$1/3. 
We verified that collisions with hydrogen are unimportant
for this hot star: setting $S_{\rm H}$=0, thus neglecting 
H collisions, has no effect on the computed line profile.

\begin{table}
\caption{Atomic data used for spectrum synthesis of the \ion{He}{i}
667.8\,nm line.}\label{atom}
\begin{tabular}{cccc}
\hline
Wavelength & $\chi_{ex}$ & log gf & isotope\\ 
nm & eV & & \\
\hline
667.8154 & 21.22 & 0.329 & $^4$He \\
667.8652 & 21.22 & 0.329 & $^3$He \\
\hline
\end{tabular}
\end{table}

It would be tempting to use other \ion{He}{i} lines
to fix the helium abundance, and to fit the 667.8\,nm line
only to derive the isotopic ratio. 
It is, however, well known that this line is discrepant with
the other He lines, both in LTE and non-LTE
\citep{WH85,SHL1,SHL2}. Thus the only reasonable approach is 
to fit the 667.8\,nm line with both the He abundance and the 
isotopic ratio as free parameters.

In Fig.\ref{He_fitiso} we show the observed line and the 
best fit both in LTE and non-LTE. 
The derived $^3$He/$^4$He ratios are 22\% and 28\%
respectively; the estimated error of this ratio is 2\%.

\section{The isotopic ratio of lithium\label{lithium}}

Lithium is an interesting element, that,
according to the standard Big Bang nucleosynthesis (SBBN),
is produced in the primordial Universe. It is expected that 
mainly $^7$Li is produced ($^6$Li/$^7$Li is predicted to be
of the order of $10^{-5}$). 
Because Li is not supposed to be produced in core-burning
nucleosynthesis, it is expected that the Li abundance in old
metal-poor stars is closely related to the Li abundance produced
by SBBN. In fact, the Li abundance is essentially the same
in all metal-poor dwarf stars, irrespective of their temperature 
and metallicity (the so-called ``Spite plateau'', \citealt{spite82}).
There seems to be now a general consensus that these metal-poor stars - as expected from
theory - have no 6Li
\citep{cayrel07,mst12,lind13}.

The lithium feature at 670.7\,nm is a doublet.
The line shift between the $^6$Li and the $^7$Li doublet
is about 7\,\kms. The main complication in the derivation
of the isotopic ratio is the non negligible thermal broadening of
the lines in the stellar atmosphere. For example, for a temperature 
of 5500\,K the full width at half maximum of the Doppler profile is 
FWHM=6.0\,\kms\ for $^7$Li and FWHM=6.5\,\kms\ for $^6$Li 
\citep[cf.][]{cayrel07}. Thus the doublet is unresolved, and the 
presence of $^6$Li is signaled by an excess absorption on the red 
wing of the feature. 

In the past, there have been several claims of detection of $^6{\rm Li}$
\citep{smith98,cayrel99,asplund06}, but now it appears that none 
withstands more sophisticated analysis. The subtleties and possible
pitfalls of the derivation of the $^6$Li/$^7$Li ratio are described in detail
by \citet{cayrel07} and \citet{lind13} for the case of metal-poor stars.
Although these two groups use a slightly different approach,
the consistent result is now a non detection of $^6$Li.

Obviously, reliable results can only be achieved with high resolution, 
high signal-to-noise spectra. In addition, the synthetic spectra used 
for the line profile fitting must account for the intrinsic line asymmetry
that is produced by convective flows in the atmospheres of cool stars and 
can mimic the presence of $^6$Li. Such analysis requires non-LTE modelling 
of the lithium line profile, based on 3D hydrodynamical model atmospheres. 
It has been demonstrated that using 1D LTE instead of 3D non-LTE line profiles 
leads to a systematic overestimation of the $^6$Li/$^7$Li isotopic ratio
by up to 2 percentage points \citep{mst12}.

\begin{figure}
\mbox{\includegraphics[bb=28 28 580 380, angle=0, width=\linewidth]
{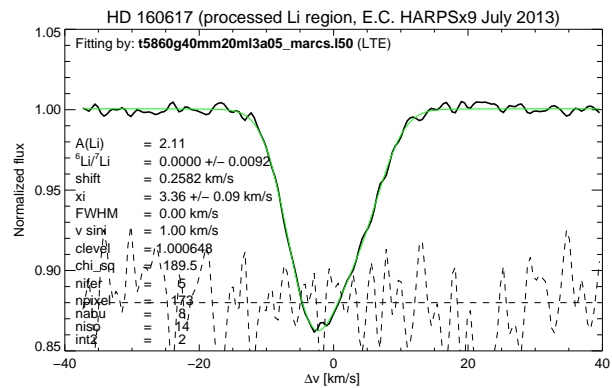}}
\caption{Best fit to the HARPS spectrum of HD\,160617 (solid black) with 
a 1D LTE synthetic spectrum (solid green), indicating a $^6$Li/$^7$Li 
isotopic ratio of $0.0000 \pm 0.0092$. The quoted uncertainty is the formal 
$1\,\sigma$ error of the $\chi^2$ fitting procedure, given the signal-to-noise 
ratio of S/N = $480$ measured on the co-added spectrum (9 sub-exposures with
a total integration time of 6.75\,h).
}
\label{hd160617a}
\end{figure}

\begin{figure}
\mbox{\includegraphics[bb=28 28 580 380, angle=0, width=\linewidth]
{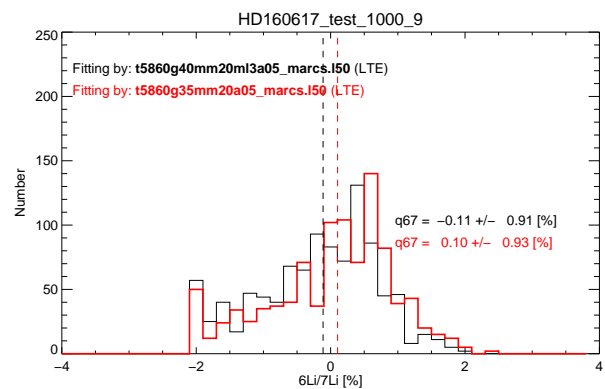}}
\caption{Probability distribution function of the $^6$Li/$^7$Li 
isotopic ratio (in \%) obtained from fitting 1000 realizations of
the spectrum of HD\,160617, generated by bootstrapping from a total of 
9 sub-exposures. Two sets of synthetic 1D LTE synthetic spectra 
were used for this experiment, assuming a gravity of $\log g = 4.0$
(black) and $\log g = 3.5$ (red), respectively. The standard deviation
of the distribution is similar in both cases ($\sigma \approx 0.009$),
and fully compatible with the formal $\chi^2$ fitting error 
(Fig.\,\ref{hd160617a}).}
\label{hd160617b}
\end{figure}

We re-analyse here the metal-poor star HD\,160617, one of
the stars for which \citet{asplund06} claimed a detection 
of $^6$Li, based on a 1D LTE analysis of their UVES spectra. 
This star (5990\,K/3.79/--1.76) has a Li abundance 
on the ``Spite plateau'', A(Li)=2.14 \citep{mst12}.
We analysed a HARPS spectrum obtained by co-adding nine exposures, 
four with 1800\,s, one with 2700\,s, and four with 3600\,s exposure 
time, observed between 2007 and 2012. The analysis on this co-added spectrum
gives a non-detection of $^6{\rm Li}$, both in 1D LTE and 3D non-LTE, 
in agreement with \citet{mst12} who used only a subset of five of the 
nine spectra analysed here.

In Fig.\,\ref{hd160617a} the 1D LTE best fit is shown, giving
$^6$Li/$^7$Li\,=\,$0.0000 \pm 0.0092$, where the quoted uncertainty is the 
formal $1\,\sigma$ error of the $\chi^2$ fitting procedure for the
 given the signal-to-noise ratio of S/N = $480$. Fitting with 3D non-LTE
synthetic spectra gives a slightly negative (unphysical) isotopic ratio,
$^6$Li/$^7$Li\,=\,$-0.0068 \pm 0.0097$.

To obtain an independent estimate of the error of $^6{\rm Li}/^7{\rm Li}$ 
due to the finite S/N ratio, we analysed a total of 1000 possible
realizations of the co-added observed spectrum, 
each constructed as a sum of nine sub-spectra, drawn randomly (allowing
repetitions) from the available set of nine different exposures (so-called
bootstrapping). The resulting distribution of the 1000 isotopic
ratio determinations is presented as a histogram in Fig.\,\ref{hd160617b}. 
The standard deviation of the distribution is about 0.009, practically
the same as the formal $1\,\sigma$ error of the $\chi^2$ fitting, suggesting 
that the latter gives a reasonable estimate of the uncertainty due to
the statistical fluctuations in the observed spectra.

\begin{figure}
\includegraphics[angle=0, width=\linewidth]{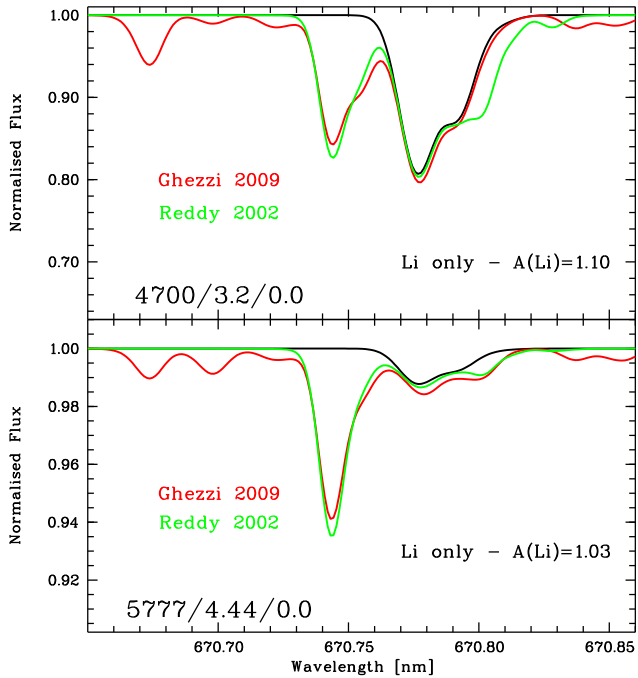}
\caption{The 1D ATLAS9 / SYNTHE synthetic spectrum in the range of the Li 
doublet at 670\,nm, based on the line list from \citet{reddy02} (solid green) 
and \citet{ghezzi09} (solid red), respectively. 
The contribution of the $^7$Li doublet (solid black) is much stronger for 
the cool solar-metallicity subgiant (upper panel) than for the Sun 
(lower panel). The blend line near 670.80\,nm in the Reddy et al.\ list 
is due to \ion{Ti}{i}.
}
\label{lisint}
\end{figure}

Finally, we point out that, in the case of metal-poor stars, the Li doublet 
is an isolated feature, and no care has to be taken of possible blending lines.
This is no longer true for solar-metallicity stars, and much care has to 
be devoted to the list of the (metal and molecular) lines falling in the Li
spectral range. Unfortunately, the blending lines are not known to
the desired  precision for an accurate derivation of the Li isotopic ratio. 
In Fig.\,\ref{lisint}, we show the synthetic spectrum of the wavelength 
range around 670.7\,nm for the case of the Sun and a solar-metallicity 
subgiant, based on two different line-lists for the blending lines,
as published by \citet{reddy02} and \citet{ghezzi09}, respectively,
Obviously, the derived $^6$Li/$^7$Li isotopic ratio depends critically on
the adopted line list, paricularly for the subgiant.

\section{Isotopes of carbon}\label{carbon}

Carbon is the fourth most abundant element in the Universe,
after H, He, and O.
Its importance is amplified because it is much easier to detect
carbon lines in stellar spectra than oxygen lines, even if oxygen
is about two times more abundant than carbon. The
main reason for this circumstance are molecular bands of CH 
(the G-band around 430\,nm) and CN (e.g. at 388\,nm), easily 
detectable in stellar spectra.
Another reason making C such an interesting element
is the discovery, over the last ten years, of ultra 
iron-poor but carbon enhanced halo stars  
\citep[see review by][]{beers05}.

\begin{figure}
\includegraphics[width=\linewidth,clip=true]{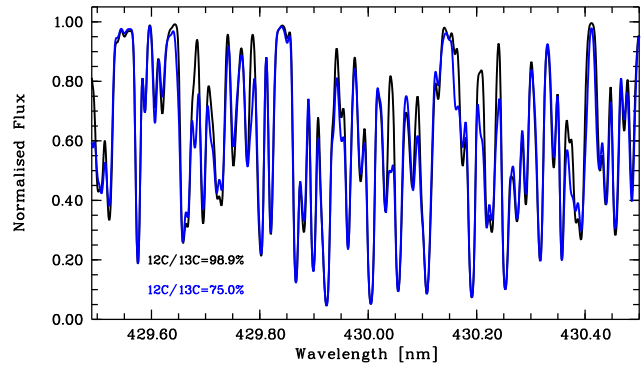}
\caption{Comparison of two synthetic spectra computed with SYNTHE from 
an ATLAS model atmosphere with parameters 4595\,K/1.1/--2.0.
The assumed isotopic ratio is
$^{12}{\rm C}/^{13}{\rm C}=98.9$\% (black), and 
$^{12}{\rm C}/^{13}{\rm C}=75.0$\% (blue), respectively.
The spectral resolution is compatible with UVES spectra
(${\rm R}\approx 80000$).}
\label{gband}
\end{figure}

\begin{figure}
\includegraphics[width=\linewidth,clip=true]{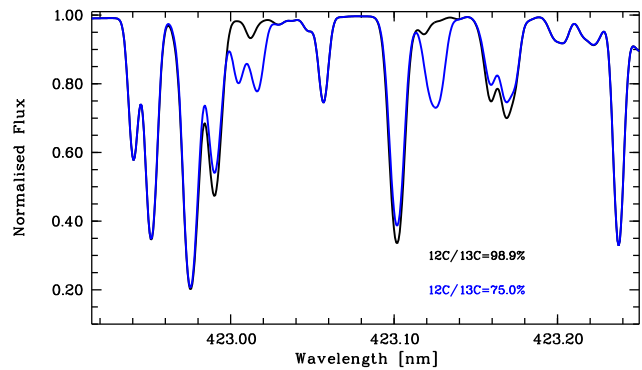}
\caption{Same as Fig.\ref{gband}, but on the tail of the G-band.
Note the isolated $^{13}{\rm CH}$ features near 423.01 and 423.12 nm.
}
\label{gbandtail}
\end{figure}

The G-band is detectable also at medium (${\rm R}\approx 10000$)
to low (${\rm R}\approx 2000$) resolving power, depending on the stellar
effective temperature and carbon abundance.
However, high spectral resolution is necessary to investigate the 
\emph{carbon isotopic ratio}.
In Fig.\,\ref{gband} the core of the G-band is shown in the case of a
metal-poor giant star for two different $^{12}{\rm C}/^{13}{\rm C}$ ratios.
In the core of the band, it is obviously very difficult to derive the 
carbon isotopic ratio. In the tail of the G-band, however, some isolated 
$^{13}{\rm CH}$ lines are accessible and allow an isotopic ratio 
determination (see Fig.\,\ref{gbandtail}).

\section{The nickel isotopic ratio in the solar photosphere}\label{nickelsec}

\begin{figure}
\includegraphics[angle=0, width=\linewidth,clip=true]{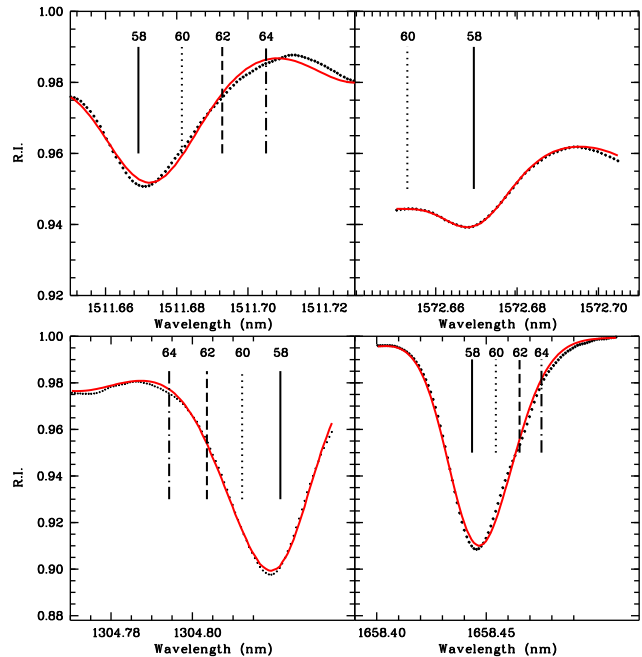}
\caption{Best fit 1D synthetic profiles (solid red) for 4 Ni lines
 super\-imposed on the disc-centre solar spectra (dotted black).}
\label{Ni_fitiso}
\end{figure}

Inspired by the work of \citet{nisun81}, we analysed the infrared 
\ion{Ni}{i} lines at 1221.6, 1304.8, 1511.6, 1572.6, 1658.4, and 1694.5\,nm,
to derive the isotopic fraction of $^{58}$Ni and $^{60}$Ni.
For this purpose, we fitted the observed disc-centre solar spectrum
by \citet{delbouille_IR}\footnote{\url{http://bass2000.obspm.fr/solar_spect.php}}.
The grid of synthetic spectra, based on the 1D empirical solar model of 
Holweger-M\"uller (\citealt{hhsunmod, hmsunmod}), has been computed with 
SYNTHE \citep{synte93,kurucz05} in its Linux version \citep{ls04,ls05}. 
The results are reported in Table\,\ref{nickel}, and the fits are shown
in Fig.\,\ref{Ni_fitiso}. Note that the 1694.5\,nm line was discarded,
because it is heavily blended. 
 
\begin{table}
\caption{Nickel isotopic fractions derived from fitting the solar infrared
Ni features, assuming fixed fractions $^{61}$Ni=1.13\,\%, 
$^{62}$Ni=3.59\,\%, $^{64}$Ni=0.91\,\%.}
\label{nickel}
\begin{tabular}{ccccccc}
\hline\noalign{\smallskip}
Wavelength & $\log gf \times$ A(Ni) & $^{58}$Ni & $^{60}$Ni & 
                                      $^{58}$Ni & $^{60}$Ni \\ 
nm &  & \% & \% & \% & \% \\
\hline\noalign{\smallskip}
   &  & \multicolumn{2}{c}{1D LTE} & \multicolumn{2}{c}{3D LTE} \\
\hline\noalign{\smallskip}
1221.6 & $-0.599\times 6.33$ & 68.3 & 26.1  & -    & -   \\
1304.8 & $-1.234\times 6.37$ & 67.9 & 26.5  & -    & -   \\
1511.6 & $-0.116\times 5.53$ & 68.3 & 26.1  & -    & -   \\
1572.6 & $-1.100\times 6.24$ & 74.8 & 19.6  & -    & -   \\
1658.4 & $-0.780\times 6.47$ & 87.7 &  6.7  & 88.5 & 5.9 \\
1694.5 & $-1.417\times 6.92$ &  -   &  -    & -    & -   \\
\hline\noalign{\smallskip}
\multicolumn{2}{c}{meteoritic$^{\,1)}$} & 68.1 & 26.2  & \multicolumn{2}{c}{}\\
\hline
\end{tabular}
Notes: $^{\,1)}$ \citet{lodders09}
\end{table}

The line at 1658\,nm is the cleanest, but yields the highest $^{58}$Ni
fraction. The 1D fit of the red wing is not quite satisfactory.
Also, a feature on the red wing suggests an unknown blend,
possibly explaining the exceedingly high (super-meteoritic) $^{58}$Ni
fraction derived from this line.

\begin{figure}
\includegraphics[angle=0, width=\linewidth,clip=true]{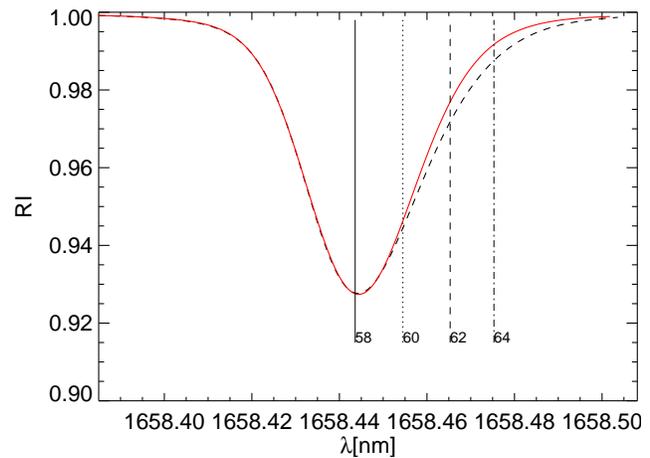}
\caption{Comparison of the 3D profile (dashed black) with the 1D profile 
(solid red) of the \ion{Ni}{i} line at 1658.4\,nm, scaled and shifted to 
match the blue wing. In both cases, $^{58}$Ni\,=\,68,1\,\%, 
$^{60}$Ni\,=\,26.2\,\% (meteoritic). The larger asymmetry of the 3D 
profile is evident.}
\label{teoprof}
\end{figure}

\begin{figure}
\includegraphics[angle=0, width=\linewidth,clip=true]{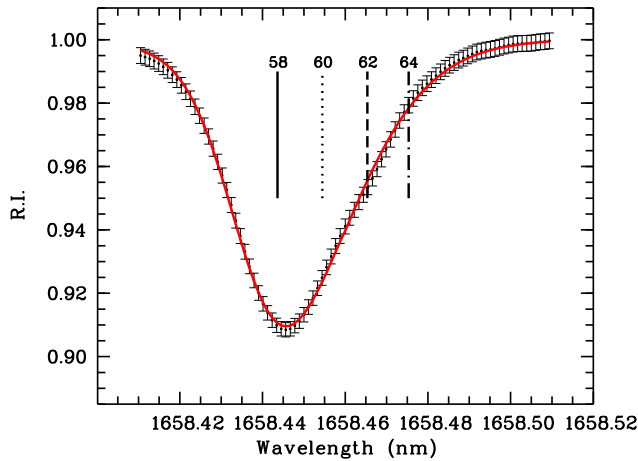}
\caption{The 3D best fit (solid red) of the Ni line at 1658\,nm superimposed 
on the disc-centre solar spectrum (black symbols).}
\label{fitiso3d}
\end{figure}

For comparison we computed with \linfor\ a grid of synthetic spectra
based on the 3D solar \cobold\ model \citep{freytag02,freytag12} 
used already in previous solar abundance analyses, e.g.\ in 
\citet{znan,2011SoPh}.
The synthetic profile based on the hydrodynamical model is composed of 
the asymmetric contributions from each of the isotopic components. The
result is a feature with an evident asymmetry, the red wing of the 3D
profile being stronger than in the 1D profile (see Fig.\,\ref{teoprof}). 
Figure \ref{fitiso3d} demonstrates that the 3D profile fits the observed 
spectrum significantly better than the 1D profile.

Nevertheless, the nickel isotopic ratio obtained from the 3D fitting is  
in good agreement with the one obtained from the 1D fitting based on 
the Holweger-M\"uller model (see Tab.\,\ref{nickel}). This is because
the convective line asymmetry affects mainly the outer wing of the
profile (see Fig.\,\ref{teoprof}), leaving the shape of the line core 
essentially unchanged with respect to the 1D case. Since the $^{58}$Ni 
to $^{60}$Ni isotopic ratio influences mainly the line core, due to the 
small isotopic shift, the derived isotopic fractions are almost insensitive 
to 3D effects.

\section{Conclusions}

The knowledge of the isotopic ratios in the chemical composition
of a star gives a much deeper insight on the underlying nucleosynthetic
processes than just the chemical abundance pattern. However,
the analysis of isotopic ratios is demanding both from the observational
and theoretical point of view. Good quality data are essential
(high-resolution and high signal-to-noise ratio).
A precise knowledge of the point source function (PSF) of the spectrograph
would certainly improve the analysis, since the shape of the line is crucial 
in deriving the isotopic ratio for several elements.
A laser frequency comb, such as the one tested on the HARPS spectrograph 
\citep{Wilken:2010:L16,Wilken}, would be the ideal tool for measuring the
PSF at any desired wavelength. For the HARPS spectrograph, the  investigation 
of the instrumental profile using the laser frequency comb is still ongoing.

In the case of $^6{\rm Li}/^7{\rm Li}$ a resolving power 
of about 100\,000 is necessary with a S/N of at least 600.
Only few spectrographs (UVES@VLT, HARPS, HIRES@Keck) can currently provide
these quality spectra, especially for metal-poor stars that are usually faint.
In the future, we expect a breakthrough with PEPSI@LBT (see Strassmeier et
al., this volume), that will allow to reach a resolving power of 300\,000, 
with a fiber-fed echelle spectrograph mounted on a telescope with a 
11.8\,m equivalent mirror. For the southern hemisphere, ESPRESSO@VLT will 
provide spectra of similar quality with a resolving power of up to 220\,000 
(see Molaro et al., this volume).

On the modelling side, we expect the development of grids of new generation
3D model atmospheres, to be utilized for an adequate interpretation of
the observed spectra by means of detailed 3D non-LTE synthetic spectra.

\acknowledgements
EC and HGL acknowledge financial support
by the Sonderforschungsbereich SFB881 ``The Milky Way
System'' (subprojects A4 and A5) of the German Research Foundation
(DFG).
PB acknowledges support from the Programme National
de Physique Stellaire (PNPS) and the Programme National
de Cosmologie et Galaxies (PNCG) of the Institut National de Sciences
de l'Univers of CNRS.

%
 \bibliographystyle{an}
%

\end{document}